\documentclass[pdflatex,sn-mathphys-num]{sn-jnl}

\usepackage{graphicx}%
\usepackage{multirow}%
\usepackage{amsmath,amssymb,amsfonts}%
\usepackage{amsthm}%
\usepackage{mathrsfs}%
\usepackage[title]{appendix}%
\usepackage{xcolor}%
\usepackage{textcomp}%
\usepackage{manyfoot}%
\usepackage{booktabs}%
\usepackage{algorithm}%
\usepackage{algorithmicx}%
\usepackage{algpseudocode}%
\usepackage{listings}%
\usepackage{booktabs}  
\usepackage{tabularx}  

\theoremstyle{thmstyleone}%
\theoremstyle{thmstyletwo}%

\theoremstyle{thmstylethree}%

\begin{document}

\title[Apple Vision Pro: Comments in Healthcare]{Apple Vision Pro: Comments in Healthcare}


\author*[1]{\fnm{Ezequiel} \sur{França dos Santos}}\email{ezequield@acm.org}

\author[2]{\fnm{Vanessa Dina} \sur{Palomino Castillo, MD}}\email{contactvanessaoficial@gmail.com}

\affil*[1]{\orgdiv{Ph.D. Student in Digital Games Development}, \orgname{IADE-U Creative University}, \orgaddress{\city{Lisbon}, \country{Portugal}}}

\affil[2]{\orgdiv{Family Physician}, \orgaddress{\city{Sao Paulo}, \country{Brazil}}}


\abstract{
This paper objectively analyzes the emerging discourse surrounding the application of Apple Vision Pro in healthcare and medical education. Released in June 2023, Apple Vision Pro represents a significant advancement in spatial computing by combining augmented and virtual reality to create new possibilities for digital interaction. We aim to compile and present recent scholarly articles, utilizing PubMed, IEEE Xplore, Google Scholar, and JSTOR as our primary sources. Non-academic publications were excluded. The research yielded six commentaries, including one pre-print and one case report, primarily expressing optimism about the technology, albeit with some concerns regarding VR/AR sickness. We highlight the need for ongoing exploration of Apple Vision Pro's capabilities and limitations and anticipate that expert opinions will further enrich this discourse.
}


\maketitle
\section{Introduction}

\subsection{Context Setting}
In the evolving landscape of technology, spatial computing has gradually merged digital elements with the physical world. One of the recent developments in this domain is Apple Vision Pro, which was introduced in June 2023\cite{appleIntroducingApple}. This technology, blending augmented and virtual reality, represents an incremental advancement in the field, offering new possibilities for user interaction with digital information.

\subsection{Purpose of the Discussion}
This paper focuses on examining the discourse surrounding the Apple Vision Pro, especially in the context of healthcare and medical education. We aim to compile and present recent commentaries' varied opinions and perspectives. We will provide a straightforward summary of how this technology might be beneficial or challenging in these particular fields, striving to maintain an objective and clear presentation of the discussions.

\section{Methodology}

We conducted a comprehensive literature review using academic databases such as PubMed, IEEE Xplore, Google Scholar, and JSTOR. The keywords "Apple Vision Pro" and "Apple visionOS" were employed to identify relevant publications.

\subsection{Criteria for Inclusion and Exclusion of Papers}
Inclusion criteria:
\begin{itemize}
    \item Publications addressing Apple Vision Pro or Apple VisionOS.
    \item Papers published after June 2023 to ensure up-to-date relevance.
    \item Works presenting empirical data, theoretical analysis, or significant commentary on the technology.
\end{itemize}
Exclusion criteria:
\begin{itemize}
    \item Papers not directly focusing on Apple Vision Pro or Apple visionOS.
    \item Non-academic articles and unrelated editorials.
\end{itemize}

\subsection{Results}
Our search yielded six papers that provided diverse perspectives on Apple Vision Pro, including comment articles, one pre-print from arXiv, and one case report. The inclusion of the pre-print was justified by its authors' relevant academic expertise and insights into the applications of Apple Vision Pro, enriching the discourse on its potential impact in healthcare and medical education.

\section{Expected Applications of Apple Vision Pro in Healthcare and Medical Education}
 The key points from each commentary are summarized in Table \ref{tab:paper_summary}, offering a concise overview of the perspectives on Apple Vision Pro's applications in these critical fields.

\subsection{Healthcare: Precision Medicine and Clinical Procedures}
\textit{Egger et al. (2023): Apple Vision Pro for Healthcare: "The Ultimate Display"? -- Entering the Wonderland of Precision Medicine.}\nocite{Egger2023}  
This paper, primarily a commentary, examines Apple Vision Pro's potential to revolutionize healthcare, particularly in precision medicine. The authors detail the device's capabilities, such as its Mixed Reality (MR) headset functionality, which combines Virtual Reality (VR) and Augmented Reality (AR) through Video See-Through (VST) technology. They discuss its applications in enhancing clinical procedures, surgical navigation, and patient care. The paper posits that Apple Vision Pro could enable clinicians to overcome certain limitations of AR in medicine, such as improving patient interaction and reducing procedural errors. They mention the problem's side effects, such as VR/AR sickness, and how the improvements in computing processing and custom lenses could reduce or solve that.

\subsection{Medical Education and Training}
\textit{Waisberg et al. (2023, Irish Journal of Medical Science): Apple Vision Pro and why extended reality will revolutionize the future of medicine.}\nocite{Waisberg2023}
In medical education, this paper, primarily a commentary, explores how Apple Vision Pro's extended reality environment could enhance learning and training for medical professionals. The authors discuss its potential to provide interactive, immersive educational experiences, such as surgical simulations and interactive learning modules, which could significantly improve medical education and training methodologies. This article does not comment on VR/AR sickness.

\subsection{Ophthalmology and Vision Science}
\textit{Waisberg et al. (2023, Eye): The future of ophthalmology and vision science with the Apple Vision Pro.}\nocite{Waisberg2023ophthalmology}
This study, primarily a commentary, delves into the transformative potential of Apple Vision Pro in ophthalmology and vision science. It discusses how the device's advanced eye-tracking and high-resolution displays can aid in precise vision screening and diagnostics. The authors envision applications in patient education and immersive diagnostics, suggesting that Apple Vision Pro could significantly impact how eye care professionals diagnose and treat vision-related conditions. This article does not comment on VR/AR sickness.

\subsection{Enhancing Accessibility in Healthcare}
\textit{Masalkhi et al. (2023, Annals of Biomedical Engineering)\nocite{Masalkhi2023}: Apple Vision Pro for Ophthalmology and Medicine.}
Masalkhi and colleagues explore, primarily as a commentary, the potential of Apple Vision Pro in enhancing accessibility, particularly for individuals with visual impairments. The paper explores the device's technical capabilities, particularly its superior display features. It discusses how these features can enhance visual experiences and accessibility in healthcare settings. The authors envision the Apple Vision Pro as a tool that could offer groundbreaking solutions in vision screening and assistance, opening new possibilities for inclusive care in ophthalmology. This article does not comment on VR/AR sickness.

\subsection{Psycological Research}
\textit{Barrett et al. (2023, Behaviour and Information Technology)\nocite{Barrett2023}: Exploring the influence of audience familiarity on speaker anxiety and performance in virtual reality and real-life presentation contexts.}
This research, primarily a commentary, investigates the utilization of Apple Vision Pro in the context of psychological research and professional development, particularly in enhancing public speaking skills. The study highlights how virtual environments created through the device can influence performance and anxiety in public speaking. By simulating various audience types, the paper suggests that Apple Vision Pro can effectively train individuals in communication skills, offering a safe and controlled environment to practice and refine public speaking abilities. This article does not comment on VR/AR sickness.

\subsection{Advancement in Medical Education}
\textit{Waisberg et al. (2023, Canadian Medical Education Journal): Apple Vision Pro and the advancement of medical education with extended reality.}\nocite{Waisberg2023canada} 
This work, primarily a commentary, explores the transformative impact of the Apple Vision Pro on medical education. The authors discuss how extended reality, a combination of virtual and augmented reality technologies, can revolutionize traditional learning methods in the medical field. They highlight the potential for immersive and interactive experiences that enhance understanding of complex medical conditions, improve diagnostic skills, and provide realistic simulations for medical training. The paper emphasizes the role of this technology in fostering a more profound, more practical learning experience for medical students and professionals, potentially leading to improved patient care outcomes. This article does not comment on VR/AR sickness.

\subsection{Neurosurgical Planning Tool}
\textit{Olexa et al. (2024): The Apple Vision Pro as a Neurosurgical Planning Tool: A Case Report.}\cite{Olexa2024}
This case report demonstrates the practical application of the Apple Vision Pro in neurosurgery, specifically in the planning and execution of a ventriculoperitoneal shunt placement for a patient with hydrocephalus. Using the Apple Vision Pro headset, surgeons were able to visualize and interact with a 3D model of the patient's anatomy, significantly aiding in the surgical planning process. The technology allowed for an enhanced understanding of the anatomical structures involved, particularly the ventricles' size and location. Following the procedure, clinicians provided feedback via a survey, reporting that the 3D models felt realistic and the integration of the user’s real-world view through the headset felt natural, without causing eye strain or fatigue. Importantly, the majority expressed a willingness to continue using this technology for future cases. This case report highlights the potential of the Apple Vision Pro to revolutionize neurosurgical planning and anatomical visualization, offering an example of its real-world application in a clinical setting and underscoring the promise it holds for improving surgical precision and patient outcomes as the device becomes more integrated into healthcare practices.

\begin{table}[h]
\centering
\caption{Summary of Results}
\label{tab:paper_summary}
\begin{tabularx}{\textwidth}{>{\raggedright\arraybackslash}X 
                              >{\raggedright\arraybackslash}X 
                              >{\raggedright\arraybackslash}X 
                              >{\raggedright\arraybackslash}X}
\toprule
\textbf{Author} & \textbf{Area} & \textbf{Application} & \textbf{Expected Improvements} \\
\midrule
Egger et al. (2023) & Healthcare & Clinical procedures, patient care & Enhanced surgical navigation, improved patient interaction \\
\midrule
Waisberg et al. (2023, Eye) & Ophthalmology & Vision screening, diagnostics & Precise diagnostics, improved patient education \\
\midrule
Waisberg et al. (2023, Irish Journal) & Medical Education & Training for medical professionals & Interactive learning, improved training methodologies \\
\midrule
Masalkhi et al. (2023) & Healthcare Accessibility & Enhanced visual experiences for the visually impaired & Improved accessibility in healthcare settings \\
\midrule
Barrett et al. (2023) & Psychological Research & VR-based emotional and behavioral studies & Enhanced research capabilities, realistic 3D rendering \\
\midrule
Waisberg et al. (2023, Canadian Journal) & Medical Education & Advanced medical training & Immersive educational experiences, enhanced diagnostic skills \\
\midrule
Olexa et al. (2024) & Neurosurgery & Surgical planning for ventriculoperitoneal shunt placement & Enhanced precision in surgical planning, improved anatomical visualization \\
\bottomrule
\end{tabularx}
\end{table}

\section{Discussion}
\subsection{Challenges and Future Directions}
The collection of papers reviewed, including commentaries and a clinical case report \cite{Egger2023} \cite{Waisberg2023} \cite{Waisberg2023ophthalmology} \cite{Masalkhi2023} \cite{Barrett2023} \cite{Waisberg2023canada} \cite{Olexa2024}, presents a predominantly optimistic view of the potential applications of Apple Vision Pro in healthcare and medical education. However, they also shed light on various challenges that must be addressed. 

Integrating Apple Vision Pro into existing healthcare systems poses a significant challenge, as illustrated by the case report \cite{Olexa2024}, which underscores the device's potential in neurosurgical planning and execution. While the device demonstrated considerable promise in enhancing surgical precision and patient outcomes, its adoption is hindered by issues such as the need for seamless integration into hospital workflows, privacy and data security concerns, and ensuring the device's accessibility and adaptability for users in clinical settings \cite{Masalkhi2023}.

The case report by Olexa et al. (2024) \cite{Olexa2024} not only exemplifies the practical application of Apple Vision Pro in a complex neurosurgical procedure but also highlights the possibilities of use cases for continued research and development.

Moreover, the high cost of Apple Vision Pro could limit its accessibility, particularly in resource-constrained settings \cite{Egger2023}. This financial barrier might restrict the technology's widespread adoption, emphasizing the need for cost reduction strategies without compromising the device's quality and functionality.

In conclusion, while the Apple Vision Pro stands on the cusp of revolutionizing healthcare and medical education through its advanced spatial computing capabilities, realizing its full potential requires a concerted effort to address these challenges. The insights provided by Olexa et al. (2024) offer valuable perspectives for future endeavors aimed at integrating this innovative technology into mainstream medical practice.

\subsection{VR/AR Side Effects}
An essential aspect missing in most commentaries is the discussion on VR/AR side effects, such as virtual reality sickness, which can manifest as nausea, disorientation, and headaches. While these technologies promise revolutionary changes, addressing their potential impact on user health and comfort is crucial, especially in prolonged or intense usage scenarios. However, the battery capacity does not allow extended usage periods since they endure more or less two hours\cite{Egger2023}.  

The pre-print article from arXiv\cite{Egger2023} provides some insights into these issues, but there is a clear need for more comprehensive research regarding Apple Vision Pro use cases.

\subsection{Apple Vision Pro Health Advisory}
In a report dated January 19, 2024, Apple Inc. issued a health advisory for users of Apple Vision Pro, particularly those with certain medical conditions such as migraines or chronic headaches, dizziness or vertigo, eye or vision conditions, such as binocular vision conditions, psychological conditions, inner ear conditions, history of dry eyes, itchiness, or swelling of the eyelids, skin allergies or sensitivities, seizures, balance or gait conditions\cite{apple2024health}. The company recommends these individuals consult with a healthcare provider before using the device. Also, updated acknolegement to when to stop using Apple Vision Pro: if you experience symptoms related to a medical condition; if your medical provider confirmed that it's safe for you to use Apple Vision Pro, but you experience severe or persistent physical discomfort, motion sickness, visual discomfort, skin irritation, or psychological symptoms; if you notice swelling, itchiness, skin irritation, or other skin reactions while using your device or after using the device\cite{apple2024health}. Always, if symptoms persist, consult with your medical provider\cite{apple2024health}.

This advisory highlights the importance of considering individual health conditions when using advanced technological devices in medical settings.

\section{Final Thoughts}
The potential applications of Apple Vision Pro are significant and promising. Besides the lack of formal publications due to waiting for the Apple Vision Pro release, the experts' comments envision an optimistic scenario.

Future research should explore the technology's benefits and drawbacks, including ergonomic concerns and potential health effects like VR/AR sickness. Understanding these dimensions will be pivotal in developing innovative and considerate solutions for user well-being.

As we navigate the advancements in spatial computing, we must keep a balanced view, acknowledging the opportunities and challenges presented by technologies like Apple Vision Pro. More research and examination will guide the technology toward a path that maximizes benefits while minimizing risks.

\section*{Acknowledgment}
The authors declare that there is no conflict of interest regarding the publication of this paper. This research did not receive any specific grant from funding agencies in the public, commercial, or not-for-profit sectors.

\bibstyle{ieee}
\bibliography{export}

\end{document}